\documentclass[aps,pra,reprint,amsmath,amssymb,longbibliography]{revtex4-2}

\usepackage{graphicx}
\usepackage{bm,color}
\usepackage{dcolumn}
\usepackage[utf8]{inputenc}
\usepackage{float}

\usepackage{array,mathtools,amssymb,booktabs}
\newcolumntype{C}{>{$}c<{$}}
\AtBeginDocument{
\heavyrulewidth=.08em
\lightrulewidth=.05em
\cmidrulewidth=.03em
\belowrulesep=.65ex
\belowbottomsep=0pt
\aboverulesep=.4ex
\abovetopsep=0pt
\cmidrulesep=\doublerulesep
\cmidrulekern=.5em
\defaultaddspace=.5em
}

\usepackage{hyperref}
\usepackage{soul}
\begin{document}

\title{Excitation of the $^{229}$Th nucleus by the hole in the inner electronic shells}

\author{M.~G.~Kozlov$^{1,2}$}
\author{A.~V.~Oleynichenko$^{1}$}
\author{D.~Budker$^{3,4}$}
\author{D.~A.~Glazov$^5$}
\author{Y.~V.~Lomachuk$^{1}$}
\author{V.~M.~Shabaev$^{6,1}$}
\author{A.~V.~Titov$^{1}$}
\author{I.~I.~Tupitsyn$^6$}
\author{A.~V.~Volotka$^5$}

\affiliation{$^1$ Petersburg Nuclear Physics Institute of NRC ``Kurchatov Institute'', Gatchina, Leningrad District 188300, Russia \\
$^2$ St. Petersburg Electrotechnical University LETI, St. Petersburg 197376, Russia\\
$^3$ Helmholtz-Institut, GSI Helmholtzzentrum fur Schwerionenforschung, Mainz 55128, Germany\\
$^4$ Johannes Gutenberg University, Mainz 55128, Germany\\
$^5$ School of Physics and Engineering, ITMO University, Kronverkskiy 49, 197101 St. Petersburg, Russia\\
$^6$ Department of Physics, St.\ Petersburg State University, Ulianovskaya 1, Petrodvorets, 198504 St.\ Petersburg, Russia}

\begin{abstract}
The $^{229}$Th nucleus has a long-lived isomeric state $A^*$ at 8.338(24)\,eV [Kraemer et al, Nature, \textbf{617}, 706 (2023)]. This state is connected to the ground state by an M1 transition. For a hydrogenlike Th ion in the $1s$ state the hyperfine structure splitting is about 0.7\,eV. This means that the hyperfine interaction can mix the nuclear ground state with the isomeric state with a mixing coefficient $\beta$ about 0.03. If the electron is suddenly removed from this system, the nucleus will be left in the mixed state. The probability to find the nucleus in the isomeric state $A^*$ is equal to $\beta^2\sim 10^{-3}$. For the $2s$ state the effect is roughly two orders of magnitude smaller. An atom with a hole in the $1s$ or $2s$ shell is similar to the hydrogenlike atom, only the hole has a short lifetime $\tau$. After the hole is filled, there is a non-zero probability to find the nucleus in the $A^*$ state. Estimates of this probability are presented along with a discussion of possible experiments on Th-doped xenotime-type orthophosphate crystals and other broad band gap materials. 
\end{abstract}

\date{\today}

\maketitle

\section{Introduction}

The isomeric state $A^*$ $(\tfrac32^+)$ of the $^{229}$Th nucleus lies $\Delta\approx 8.3$\,eV above the ground state $A$ $(I^\pi=\tfrac52^+)$ \cite{Seiferle2019,Sikorsky2020,Kraemer2023}. This state is widely considered as ideal candidate for the creation of a nuclear optical clock \cite{PT03,Rellergert2010}. That, as expected, will further stimulate many practical and fundamental applications, including studying new physics \cite{Fla06c,BDFP09,SBDJ18,FBF20}, observing Dicke and M\"ossbauer effects in the UV range \cite{Dicke_53,Tkalya2011}, and others. The properties of this metastable state have been investigated experimentally \cite{WSLN16,TOGM18,SPKT18} but these studies are hampered due to the limited quantities of isomeric nuclei. That is why lot of attention is paid to find an optimal way to populate isomeric state.

Thorium-229 in the $A^*$ state is usually 
produced in nuclear reactions (for example, in $\alpha$-decay of $^{233}$U). Alternatively, one can try to excite $^{229}$Th to the isomeric state \cite{Tkalya1992,Karpeshin1998,Tkalya2020,Beeks2022,JBK23,masuda2019x}. Here we discuss a relatively simple way to do this using the hyperfine mixing of nuclear states \cite{LOP66,WyZy93,KBT99,Tkalya2016a,SGR22}.
This may help to perform accurate measurement of the nuclear transition frequency, which is crucial for making an optical nuclear clock, see recent review \cite{TKMS24}.

The hyperfine structure constants for the 
hydrogenlike thorium are $a(1s)\approx 0.68$\,eV and $a(2s)\approx 0.09$\,eV \cite{SGR22}. The off-diagonal hyperfine matrix element $b$ between nuclear states $A$ and $A^*$ is expected to be of the same order of magnitude. 
Following Refs.\,\cite{Minkov2019,Shigekawa2021,SGR22}, we assume that $|b(1s)|\approx 0.24$ eV and $|b(2s)|\approx 0.03$ eV. The hole in the $1s$, or $2s$ shell has the hydrogenlike wavefunction and, therefore, the same values of the hyperfine parameters. 

Consider a process of the hole production in $1s$ or $2s$ shell of neutral thorium by photoionization \cite{Puri1999,AjayKumar_2001,Kumar2002}, or by electron-impact ionization \cite{XuXu95,rahangdale2015absolute}. The hole lives for a time $\tau$, and during this time the non-diagonal hyperfine interaction mixes two lowest nuclear states. When the hole is filled, the nucleus is left in one of these two states. Below we estimate the probability to find the nucleus in the $A^*$ state. 
The Dirac-Fock energies are $\varepsilon_{1s} \approx 111$\,keV and $\varepsilon_{2s} \approx 21$\,keV. The lifetime of the hole is determined by the radiative width \cite{CaPa01}:
\begin{align}
    \label{eq:est}
    &\Gamma_{1s}\approx 90\,\mathrm{eV}; &\Gamma_{2s}\approx 14\,\mathrm{eV}. 
\end{align}

Suppose, we use a high-energy photon to excite an electron to the continuum ($E_\gamma\gtrsim 100$\,keV for the K-shell and $E_\gamma\gtrsim 20$\,keV for the L-shell). The interaction operator has the form $j_\mu A^\mu$, where $A^\mu$ is vector potential and the current, $j_\mu=j_\mu^{(e)}+j_\mu^{(N)}$, includes electronic and nuclear parts. However, the nucleus has the $A^*$ level at 8.3\,eV and only a few levels in the sub-MeV range, the first one being at 29 keV. Thus, for the photon energies far from the nuclear resonances we can neglect the nuclear term. We conclude that photons interact only with the electronic degrees of freedom and do not affect the nucleus. In the shake-off approximation, the electron disappears instantaneously, leaving the nucleus unchanged. For closed electronic shells there is no magnetic interaction with the nucleus and the nucleus remains in its ground state. When the hole is created, the nucleus starts to interact magnetically with the spin of the hole. The non-diagonal part of this interaction leads to mixing between nuclear states $A$ and $A^*$. The hole lives for an average time $\tau$ before it is filled with an electron from one of the higher shells. Once the hole is refilled, magnetic interaction ``turns off'' and the nucleus is left in one of its eigenstates. Naively one can expect that the probability $P^*$ to find the nucleus in the state $A^*$ is on the order of $(b\tau)^2$. Since this probability depends on both $b$ and $\tau$, it may be beneficial to have a hole in a shell other than the K shell. Below we estimate $P^*$ more accurately, and find that it is almost the same for K and L$_1$ shells.
 
 \begin{figure}[ht]
  \includegraphics[scale=1.0]{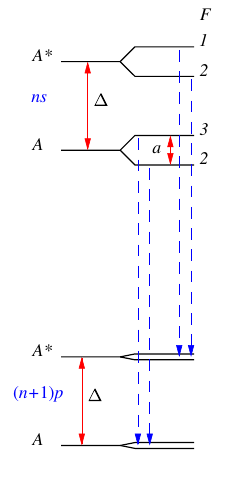}
  \caption{Schematic level structure of $^{229}$Th ion with a hole in the inner $ns$ shell, $n=1,2$. $\Delta\approx 8.3$~eV is the energy of the isomeric nuclear state $A^*$ $(I^\pi=\tfrac32^+)$. Blue dashed lines are allowed transitions of the hole to the $(n+1)p$ shell. The hyperfine splittings for the two nuclear states are comparable, while for the electronic $(n+1)p$ state it is much smaller than for $ns$ state. The fine structure for the $p$ shell is not shown.}
 \label{fig1}
 \end{figure}

\section{Four-level model}
A simplified level structure of the atom with a hole is shown in Fig.\,\ref{fig1}. The upper four levels correspond to the hole in the inner shell $ns$, where $n=1,2,$\dots\ The lower four levels describe the atom with a hole in the $(n+1)p$ shell. The hyperfine structure for these levels is much smaller and we neglect it. For the $ns$ hole we include the non-diagonal hyperfine interaction between two levels with the same total angular momentum $F=2$. 

Let us consider a four-level model. In addition to the two $F=2$ levels for the $ns$ shell we add two levels in the $(n+1)p$ shell, so that radiative decay $ns\to (n+1)p$ can take place. Due to the fine and hyperfine structure, there are, in fact, more than two levels where the $ns$ hole can decay, but this should not significantly affect our conclusions. The effective Hamiltonian of this system has the form:
\begin{align}\label{eq:ham4}
    H =
    \left(
    \begin{array}{cccc}
       0  &  0  &  0  &  0  \\
       0  & \Delta &  0  &  0  \\
       0  &  0  & \omega &  b  \\
       0  &  0  &  b  & \omega+\Delta 
    \end{array}
    \right) \,.
\end{align}
Here $\Delta \approx 8.3$ eV is the nuclear excitation energy, $b$ is the non-diagonal hyperfine matrix element, and $\omega$ is the energy interval to the $(n+1)p$ shell. We will describe this system by the density matrix $\rho$:
\begin{align}\label{eq:rho}
    \rho =
    \left(
    \begin{array}{cccc}
       p  &  0  &  0  &  0  \\
       0  & p_* &  0  &  0  \\
       0  &  0  &  s  &  r  \\
       0  &  0  &  r^*&  s_* 
    \end{array}
    \right) \,,
\end{align}
where $s$ and $p$ correspond to the hole in $ns$ and $(n+1)p$ shells respectively and the nucleus in the ground state $A$; $s_*$ and $p_*$ correspond to the  nucleus in the excited state $A^*$.
The density matrix $\rho$ obeys the Liouville equation,
\begin{align}
 \dot{\rho} &=-i\left[H,\rho\right] + \Lambda\,,
 \label{eq:master}
\end{align}
where we set $\hbar=1$ and the matrix $\Lambda$ describes decay and pumping of the levels: 
\begin{align}\label{eq:Lambda}
    \Lambda = \Gamma
    \left(
    \begin{array}{cccc}
         s  &  0  &  0  &  0  \\
         0  &  s_*&  0  &  0  \\
         0  &  0  & -s  & -r  \\
         0  &  0  & -r^*& -s_* 
    \end{array}
    \right) \,.
\end{align}

We look for the solution of Eq.\,\eqref{eq:master} with the initial condition $s(0)=1$ and other variables equal to zero. At the asymptotic limit $t=\infty$ only $p$ and $p_*$ are nonzero due to the decay of the $ns$ hole. In this model the probability to find the nucleus in the excited state is $P^*=p_*(\infty)$.

Plugging Eqs.\ \eqref{eq:ham4} and \eqref{eq:Lambda} in \eqref{eq:master} we get: 
\begin{align}\label{eq:system}
\left\{
\begin{array}{lcl}
   \dot{p}  & = & +\Gamma s \,, \\
   \dot{p}_*  & = & +\Gamma s_* \,,  \\
   \dot{s}  & = & -ib(r^*-r)-\Gamma s \,, \\
   \dot{s_*}  & = & +ib(r^*-r)-\Gamma s_* \,, \\
   \dot{r}  & = & +ib(s-s_*)+i\Delta r-\Gamma r \,, \\
   \dot{r^*}  & = & -ib(s-s_*)-i\Delta r^*-\Gamma r^* \,. \\
\end{array}
\right.
\end{align}
Remaining parameters of the density matrix are not linked to the parameter $s$ and, therefore, are all equal to zero at all times. This argument justifies the form of the density matrix assumed in Eq.\,\eqref{eq:rho}.

We see that the variables $s$, $s_*$, $r$, and $r^*$ do not depend on $p$ and $p_*$. We can make the following substitution:
\begin{align}\label{eq:subst}
\left\{
\begin{array}{lcl}
   s & = & \frac{1}{2}e^{-\Gamma t} \left(1+\zeta\right) \,, \\
   s_* & = & \frac{1}{2}e^{-\Gamma t} \left(1-\zeta\right) \,, \\
   r & = & e^{-\Gamma t} \left( \xi+i\varphi\right) \,. \\
\end{array}
\right.
\end{align}
After that, the four lower equations of system \eqref{eq:system} give:
\begin{subequations}
\label{eq:zpx}
\begin{align}
    \label{eq:zeta}
    \dot{\zeta} &= -4b\varphi\,, &\zeta(0) = 1 \,, \\
    \label{eq:phi}
    \dot{\varphi} &= b\zeta +\Delta \xi\,, &\varphi(0) = 0 \,, \\
    \label{eq:xi}
    \dot{\xi} &= -\Delta \varphi\,, &\xi(0) = 0 \,.
\end{align}
\end{subequations}
Let us differentiate \eqref{eq:phi} and then use \eqref{eq:zeta} and \eqref{eq:xi} for the right hand side:
\begin{align}
    \ddot{\varphi} &= b\dot{\zeta} +\Delta\dot{\xi} = -(4b^2+\Delta^2)\varphi \,,
    \quad\Rightarrow 
    \nonumber \\
    \varphi&=\varphi_0\sin{\sqrt{4b^2+\Delta^2}t} \,.
\end{align}

Equations (\ref{eq:zeta}, \ref{eq:xi}) show that $\zeta$ and $\xi$ oscillate at the same frequency as $\varphi$ and have the phase shift $\pi/2$. The solution, which satisfies initial conditions has the form:
\begin{align}
    \zeta &= (1-A) + A\cos{\sqrt{4b^2+\Delta^2}t} \,, \\
    \xi &= B \left(1-\cos{\sqrt{4b^2+\Delta^2}t}\right) \,.
\end{align}
Substituting these expressions into \eqref{eq:zpx}, we get:
\begin{align}
    \label{eq:A}
    &A = \frac{4b^2}{4b^2+\Delta^2} \,,
    &B = -\frac{b\Delta}{4b^2+\Delta^2} \,,
    \\
    \label{eq:phi0}
    &\varphi_0 = \frac{b}{\sqrt{4b^2+\Delta^2}} \,.
\end{align}

We can now write the required matrix element $s_*$:
\begin{align}
    s_* &= e^{-\Gamma t} 
    \frac{2b^2}{4b^2+\Delta^2}
    \left(1-\cos{\sqrt{4b^2+\Delta^2}t}\right) \,.
\end{align}
Substituting $s_*$ into second Eq.\ \eqref{eq:system} and integrating it from zero to infinity, we find:
\begin{align}\label{eq:prob4}
    P^*= p_*(\infty) &= \frac{2b^2}{\Gamma^2+4b^2+\Delta^2}
    \approx \frac{2b^2}{\Gamma^2+\Delta^2}\,,
\end{align}
where we took into account that $|b|\ll \Delta$. Note, that neglecting $\Delta$, we reproduce the estimate given above up to a factor of 2.
Numerically Eq.\ \eqref{eq:prob4} gives:
\begin{align}\label{eq:P_1s4}
    &P^*(1s) \sim 1\cdot 10^{-5}\,,
    &P^*(2s) \sim 7\cdot 10^{-6}
    \,.
\end{align}
These values are much smaller than what was recently obtained by \citet{Kar23} using the expression for the hydrogen-like ion \cite{KBT99}, where electronic levels have negligible widths. By putting $\Gamma=0$ in \eqref{eq:prob4} we get values of the same order of magnitude as in that work.  

We see that the probability to excite the nucleus for the $1s$ and $2s$ holes is almost the same. The numerator in~\eqref{eq:prob4} rapidly decreases for the principle quantum numbers $n>2$, while the denominator remains practically constant. Thus, for the higher shells the probability to excite the nucleus is much smaller. Taking into account much larger energy of the K hole, we conclude that the optimal variant is to use $2s$ holes.  

\section{Yield estimate of the isomeric nuclei} 

Keeping in mind the narrow width of the nuclear transition, its experimental study requires a large number of nuclei in the state $A^*$. The yield of these metastable nuclei depends on the probability $P^*$, the number of the $^{229}$Th ions in the sample $N$, and the photon flux $F_p$ from the X-ray source, if we use photo-ionization, or the electron flux $F_e$ if we use electron impact to produce $2s$ holes. For the H-like ions in the $1s$ state the level width is negligible, while K hole has the width about 100 eV. Thus, the probability to excite the nucleus in the H-like ion is almost 100 times larger. However, this probability is multiplied by the number $N$ of $^{229}$Th nuclei in the sample. In a solid target $N$ is proportional to the Avogadro number, while in the beam of the H-like ions it is many orders of magnitude smaller. 

To have large $N$ we need a solid sample with high concentration of $^{229}$Th. The host crystal must be transparent for the 8.3\,eV photons, which means a sufficiently broad band gap. Moreover, the lifetime of the nuclear $A^*$ state depends on the oxidation state of thorium because the internal conversion and electron-bridge processes drastically decrease the lifetime of the isomeric state for Th and Th$^+$ \cite{Tkalya2003,KaTr07,PorFla10}. Thus, in order to suppress these processes the Th$^{2+}$, Th$^{3+}$ or Th$^{4+}$ ions immobilized in the solid matrix are desirable. To sum up, an important problem to be studied separately is finding a suitable material with such a wide band gap. Last but not least,
this material should have high radiation resistance.

In general there are two ways towards preparation of such $^{229}$Th-containing solid samples. Within the first one thorium ions (or ions of the $^{229}$Th isotope predecessors) are directly implanted into a thin film of the material. A notable experiment of this type is reported in \cite{BCKL18} where this technique was used to determine the isomeric transition energy (some signal was detected, however, characteristic energy was $7.1\,(+0.1-0.2)$~eV, which strongly disagrees with other experiments). The beam of thorium ions was obtained by laser ablation of a $^{229}$Th-enriched target and implanted in a thin film of SiO$_2$ (with band gap $E_g \sim 9$\,eV). The number of thorium nuclei in the sample was estimated to be $N \approx 3\cdot10^{12}$ \cite{BCKL18}. The other class of materials suitable for such experiments with wide band gap (10\,eV and higher) is fluorides of alkali and alkaline-earth metals. In the recent experiment~\cite{Kraemer2023} radioactive ion beams of nuclei with the mass number $A = 229$ ($^{229}$Fr, $^{229}$Ra) were implanted in CaF$_2$ (band gap of 12.1\,eV\,\cite{Rubloff1972}) and MgF$_2$ (12.4\,eV\,\cite{Thomas1973}) crystals at 30\,keV, with subsequent series of $\beta$-decays resulting in $^{229m}$Th nuclei. The observed signal from the M1 nuclear transition $A^*\to A$ suggested that only several percent of the implanted thorium ions decayed via this channel. This could be explained by the fact that low-oxidation states of thorium are favorable in such crystal environments. This is in qualitative agreement with the previous theoretical predictions \cite{Dessovic2014,Pimon2020}. The bottleneck of experiments with such beams is their relatively low intensities ($\sim10^6$ particles per second), which is not promising for obtaining $N > 10^{10}$ in the prepared sample. Moreover, fluorides are unstable under such beams and undergo metamictisation processes.

The other way is to perform experiments with crystals of pure thorium compounds or some synthetic Th-doped materials. To the best of our knowledge, ThF$_4$ is the only purely thorium compound with a wide band gap (10.4~eV~\cite{Gouder2019}); the oxidation state of Th in this compound is +4, thus preventing internal conversion processes. One can also consider the other fluoride matrix, lanthanum fluoride LaF$_3$, whose band gap was also found to be quite wide, 9.7\,eV\,\cite{Wiemhofer1990}. In this case the crystal should be doped with Th$^{3+}$ impurity ions. Band gaps for many other rare-earth fluorides are smaller and these do seem promising. The radiation resistance of these fluorides is questionable. Among other materials which can be regarded as matrices for Th ions, there are yttrium and lutetium orthophosphates (YPO$_4$ and LuPO$_4$), which seem to be the most suitable ones due to their large band gaps (8.6--9.4\,eV for YPO$_4$~\cite{Balcerzyk2000,Makhov2002,Wang:09,Poolton2010} and 8.6--9.3\,eV for LuPO$_4$~\cite{Nakazawa:77,Lempicki1993,Balcerzyk2000,Makhov2002}; one can expect higher band gap values for perfect crystals) and high chemical and radiation resistance (see \cite{Urusov:12,Cutts:16} and references therein). Natural YPO$_4$ is the xenotime mineral; it is known that xenotime does not undergo metamictisation on geological time scales. This is due to its rigid crystal lattice formed by the PO$_4$ groups which is not destroyed 
by radiation.
The oxidation state of Th in YPO$_4$ is expected to be +3, this was also confirmed theoretically~\cite{Lomachuk2020}; the same could be expected for LuPO$_4$ due to the same crystal structure and proximity of ionic radii and electronegativiy of Y$^{3+}$ and Lu$^{3+}$ cations. 
Note that chemical synthesis and subsequent growth of crystals of either ThF$_4$ or phosphates considered here are not problematic; appropriate techniques are well established.

Despite these unique properties of phosphate matrices, there are several issues still to be addressed. First of all, large amounts of thorium dopant ions and other point defects can result in a decrease in the band gap. The presence of thorium ions leads to significant lowering of symmetry of local sites~\cite{Lomachuk2020} and, at high concentrations of Th, the low-symmetry monazite-type structure is more stable, with typically small band gaps ($\sim 3$\,eV for monazite CePO$_4$~\cite{PalmaRamrez2015}). Moreover, one can also expect the appearance of Th-rich domains with monazite structure. The optimal concentration of thorium ions avoiding a significant decrease in the band gap can be estimated theoretically but more reliable values should be   obtained experimentally in real crystalline samples. Typical mole fraction of impurity ions in typical synthetic lanthanide-doped phosphate crystals is of order 0.1\%. To estimate the number of $^{229}$Th nuclei $N$ achievable in such a solid state experiment one can use the value of molality measured for the $^{229}$Th NIST Standard Reference Materials~\cite{NIST-SRM}, which was found to be $1.2\cdot 10^{-10}$\,mol\,g$^{-1}$~\cite{Essex2018}. Consider a macroscopic sample of Th-doped xenotime of mass 1\,g. It will contain $\approx 1.3 \cdot 10^{-3}$\,g of thorium. Multiplying this value by the molality from~\cite{Essex2018}, one obtains $1.5\cdot 10^{-13}$\,mol or $N \sim 10^{11}$ nuclei of $^{229}$Th for such a sample. This is comparable with $N \sim 10^{12}$ reported for the experiment with SiO$_2$ thin films~\cite{BCKL18}. However, an experiment with xenotime does not require any beams and seems to be simpler from the point of view of preparation. Moreover, $^{229}$Th can be produced in nuclear reactors~\cite{Hogle2016} and then used to prepare $^{229}$Th-rich samples of doped xenotime with $N \gg 10^{12}$.

Another non-obvious obstacle is the possible presence of additional narrow band gaps due to the $7p$-states of the substituting Th$^{3+}$ ions which in principle can overlap with the resonance frequency of nuclear isomeric transition. The $P_{1/2}^o$ and $P_{3/2}^o$ states of the Th$^{3+}$ atomic ion occur at 7.47 and 9.06\,eV, respectively~\cite{Klinkenberg:49}, but in the solid matrix they can be significantly shifted down (like energy levels of the  Ce$^{3+}$ impurity ions in cerium-doped xenotime, see~\cite{Pieterson:02}).
A theoretical study of these impurity levels are presented in a separate paper~\cite{Oleynichenko2023}.

\section{Efficiency}

Let us compare the efficiency of our method with that of Ref.\ \cite{masuda2019x}, where resonant X-ray beam in SPring-8 facility was used to excite the nucleus to the 29 keV state, which then predominantly decays to the isomeric state $A^*$. Assuming the same target and the same X-ray source we compare the production rate of the isomeric nuclei. Both methods use X-ray beam to excite thorium atom from the ground state $g$ to the intermediate state $e$, which decays to the final state $f$ where the nucleus is in the state $A^*$. The number of excited nuclei is proportional to the photon-absorption cross section, $\sigma_{ge}$, the branching ratio of the decay $e \to f$, $R_{ef}$, and the photon flux $F(\Delta)$, where $\Delta$ is the width of the spectral window used. Relative efficiency of two methods is given by the ratio:
\begin{align}\label{eq:efficiency1}
    E &= \frac{\sigma_{ge}}{\sigma^M_{ge}}
    \cdot\frac{R_{ef}}{R^M_{ef}} 
    \cdot\frac{F(\Delta)}{F(\Delta^M)}\,,
\end{align}
where the superscript $M$ corresponds to the method of \citet{masuda2019x}. 

According to \cite{Sco73,PCM95} the cross section for the photo-excitation of the $2s$ hole at 30 keV is:
$\sigma_{ge} = \sigma_{2s} = 6.83\cdot 10^3\, b$. The branching ratio follows from estimate \eqref{eq:P_1s4}: 
$R_{ef}=\tfrac{5}{12} P^*(2s) \sim 3\cdot 10^{-6}$,
where $\tfrac{5}{12}$ is the fraction of the holes with quantum number $F=2$.

To estimate the cross section of the photo-excitation of the nuclear 29 keV level we use the formula from \cite{BKD04} (see problems 3.5 and 3.6):
\begin{align}\label{eq:sigma_nuc}
    \sigma^M_{ge} = \sigma_\mathrm{nuc} 
    &\approx \frac{\lambda^2}{2\pi} \frac{\Gamma_e}{\Delta^M}
    = 4.8 \cdot 10^{-2}\, b\,,
\end{align}
where $\lambda$ is the wavelength of the photon, $\Gamma_e$ is the radiative width of the 29 keV nuclear level, and we substituted values from Ref.\ \cite{masuda2019x}: $\Gamma_e = 1.7 \cdot 10^{-9}$ eV,
$\Delta^M = 0.1$ eV, and $R^M_{ef} \approx 0.9$. Equation \eqref{eq:efficiency1} leads to the following relative efficiency:
\begin{align}\label{eq:efficiency2}
    E &\approx 0.18 \frac{F(\Delta)}{F(\Delta^M)}\,.
\end{align}

According to this estimate the $2s$ hole excitation gives about 20\% of the signal in the experiment \cite{masuda2019x}. A particular silicon crystal was used in the final experiment to filter photons within 0.1 eV window, while another crystal gave the window of 3.6 eV and 80 times higher flux. For such flux the production rate via $2s$ hole excitation is 14 times higher than what was observed in Ref.\ \cite{masuda2019x}.

We see that simply by increasing the width of the spectral window in the experiment \cite{masuda2019x} we can increase production of the isomeric nuclei. This will also increase the dissipated power on the target. For the 0.1 eV window the flux was $I=1\cdot 10^{12}$~s$^{-1}$. The total absorption cross section of the photon for thorium atom is: $\sigma_a=14\cdot 10^3\, b$ \cite{Sco73}. The fraction of the absorbed photons, $f=\sigma_a n h$, where $n$ is concentration of Th and $h$ is target thickness. The target had diameter $d=0.4$~mm, thickness $h=0.2$~mm, and contained $N=6.3\cdot 10^{14}$ thorium nuclei \cite{masuda2019x}, thus, $n=4N/(\pi d^2h)=0.25\cdot 10^{20}\, \mathrm{cm}^{-3}$ and $f=0.7\%$. The power on the target is $\hbar\omega If=0.03$~mW. For the 80 times higher flux the dissipated power is still about 3 mW. Tuning the photon energy closer to the $2s$ cross section peak at 20.65 keV we gain another factor of 2.6 in the production rate \cite{Sco73,PCM95}.


\section*{Acknowledgements}

We are grateful to Christoph D\"ullmann, Junlan Jin, Feodor Karpeshin, Thorsten Schumm, and Peter Thirolf for valuable discussions and comments. 
D.B. contributed to this work prior to February 2022. He was supported in part by the Deutsche Forschungsgemeinschaft (DFG, German Research Foundation) Project ID 390831469:  EXC 2118 (PRISMA+ Cluster of Excellence) and the Helmholtz
Excellence Network ExNet02.
The work on the analysis of possible solid state experiments by A.V.O., Y.V.L. and A.V.T. was supported by the Russian Science Foundation (Grant No. 20-13-00225, https://rscf.ru/en/project/23-13-45028/).

\bibliography{./main}

\end{document}